\documentclass{PoS}
\usepackage{aas_macros}
\usepackage{layouts}
\title{Six years of VERITAS observations of the Crab Nebula}

\ShortTitle{Six years of VERITAS observations of the Crab Nebula}
\author{\speaker{Kevin Meagher} for the VERITAS Collaboration\thanks{veritas.sao.arizona.edu}\\
  School of Physics and Center for Relativistic Astrophysics, Georgia Institute of Technology, 837 State Street NW, Atlanta, GA 30332-0430\\
  E-mail: \email{kevin.meagher@physics.gatech.edu}
  }
\abstract{The Crab Nebula is the brightest source in the very-high-energy (VHE) gamma-ray sky and one of the best studied non-thermal objects. The dominant VHE emission mechanism is believed to be inverse Compton scattering of low energy photons on relativistic electrons. While it is unclear how the electrons in the nebula are accelerated to energies of $10^{16}$ eV, it is general consensus that the ultimate source of energy is the Crab pulsar at the center of the nebula. Studying VHE gamma-ray emission provides valuable insight into the emission mechanisms and ultimately helps to understand the remaining mysteries of the Crab, for example, how the Poynting dominated energy flow is converted into a particle dominated flow of energy. We report on the results of six years of Crab observations with VERITAS comprising 115 hours of data taken between 2007 and 2013. VERITAS is an array of four 12-meter imaging air Cherenkov telescopes located in southern Arizona. We report on the energy spectrum, light curve, and a study of the VHE extension of the Crab Nebula.}

\FullConference{The 34th International Cosmic Ray Conference,\\
		30 July- 6 August, 2015\\
		The Hague, The Netherlands}

\newcommand{\citep}{\cite}
\newcommand{\farcm}{'\!\!.}
\newcommand{\fdg}{^{\circ}\!\!\!.}
\newcommand{\arcmin}{'}

\begin{document}

\section{Introduction}
The Crab Nebula and pulsar are two of the best studied astrophysical sources of non-thermal radiation in the sky. Their exceptional brightness across the electromagnetic spectrum and relative proximity to us make both objects excellent targets to study a cosmic particle-accelerator in great detail. 

The central engine of the Crab is a pulsar that provides the energy and particles to power one of the brightest pulsar wind nebulae we know. The radio emission like any other pulsed emission and particle outflow from a pulsar originates from the vicinity or from within the compact magnetosphere that co-rotates with the neutron star and, therefore, appears pulsed. At a distance of $\sim0.1\,$pc from the neutron star the pressure of the outflowing cold pulsar wind equals the pressure from the nebula and a standing reverse shock is formed\citep{ReesGunn1974,KennelCoroniti1984}. It is in the shock and downstream of it where electrons cool via synchrotron radiation that is responsible for the non-thermal emission of the nebula from radio up to hundreds of MeV\citep{Achterberg2001}. 

In this paper we present VERITAS observations of the Crab Nebula over 7 years covering observing periods between Fall 2007 and Fall 2014. We present spectral analysis, a search for the extension of the nebula, and a study on the variability.

\section{VERITAS}
\label{veritas}
VERITAS is an array of four imaging atmospheric Cherenkov telescopes designed to detect very-high-energy gamma rays between 80\,GeV and several ten TeV\citep{Holder2006,Holder02011}. The array is located at the Fred Lawrence Whipple Observatory in southern Arizona at an elevation of 1280\,m above sea level. 

The array became fully operational with all four telescopes in Fall 2007. In 2009 one of the telescopes had been relocated to its present location giving the array its present diamond shaped layout with approximately 100\,m separation between the telescopes, yielding an increase in sensitivity of about 30\% \citep{Perkins2009}. In Summer 2012 the telescope cameras had been upgraded with high efficiency photomultipliers resulting in a sub 100\,GeV energy threshold and higher sensitivity.

\section{Observations}
\label{observations}
Observations of the Crab are carried out with VERITAS on a regular basis. Besides the primary motivation of studying the physics of the nebula and pulsar the observations  also provide an easy way to monitor the stability of the instrument under the assumption that the Crab is a non-varying source. The data are also used to determine the sensitivity of the instrument, verify the Monte Carlo simulations. 

Crab observations are carried out with all observing modes of VERITAS, i.e. all zenith angles, dark time, moon light observations etc. However, the majority of the data, and this data is used in this analysis, are recorded in the Zenith angle range between 8 degrees, where the Crab culminates at the VERITAS site, and 35 degrees. 

Data analyzed for this work has been taken between Fall 2007, when all four telescopes became operational, until Summer 2012. In the analysis the data is divided into three epochs, each representing a significant different telescope configuration, requiring a new Monte Carlo simulation. Epoch one goes from Fall 2007 to Summer 2009, when one of the telescopes was relocated. Epoch two spans from Summer 2009 to Summer 2012 when all VERITAS cameras were upgraded with high-efficiency PMTs. Epoch three started in Summer 2012 and is still ongoing

The Crab is visible to VERITAS at Zenith angles less then 35 degrees from September to March, which are the months during which the analyzed data had been recorded. All observations presented in this paper were taken with a standard wobble offset of $0\fdg5$ from the source position, with the direction of the offset alternating between each of the four cardinal directions. Wobble observations allow background to be estimated from off source data in the field of view, canceling any systematic effects\citep{Aharonian2001,Berge2007}.

For the event selection \emph{standard cuts} are used that are optimal for such a strong source as the Crab. All results presented here had been produced with two different analysis packages in two independent analyses. 

\section{Spectrum}
\label{spectrum}

The very-high-energy gamma-ray spectrum of the Crab Nebula has been measured extensively and measurements are mostly limited by systematic uncertainties but at the highest energies above 10\,TeV. The biggest systematic effect in the spectral reconstruction are uncertainties in the energy reconstruction that lead to shifts in the reconstructed flux. At the lowest energies systematic uncertainties are dominated by effects of the trigger electronics. 

\begin{figure}
  \includegraphics[width=0.5\textwidth]{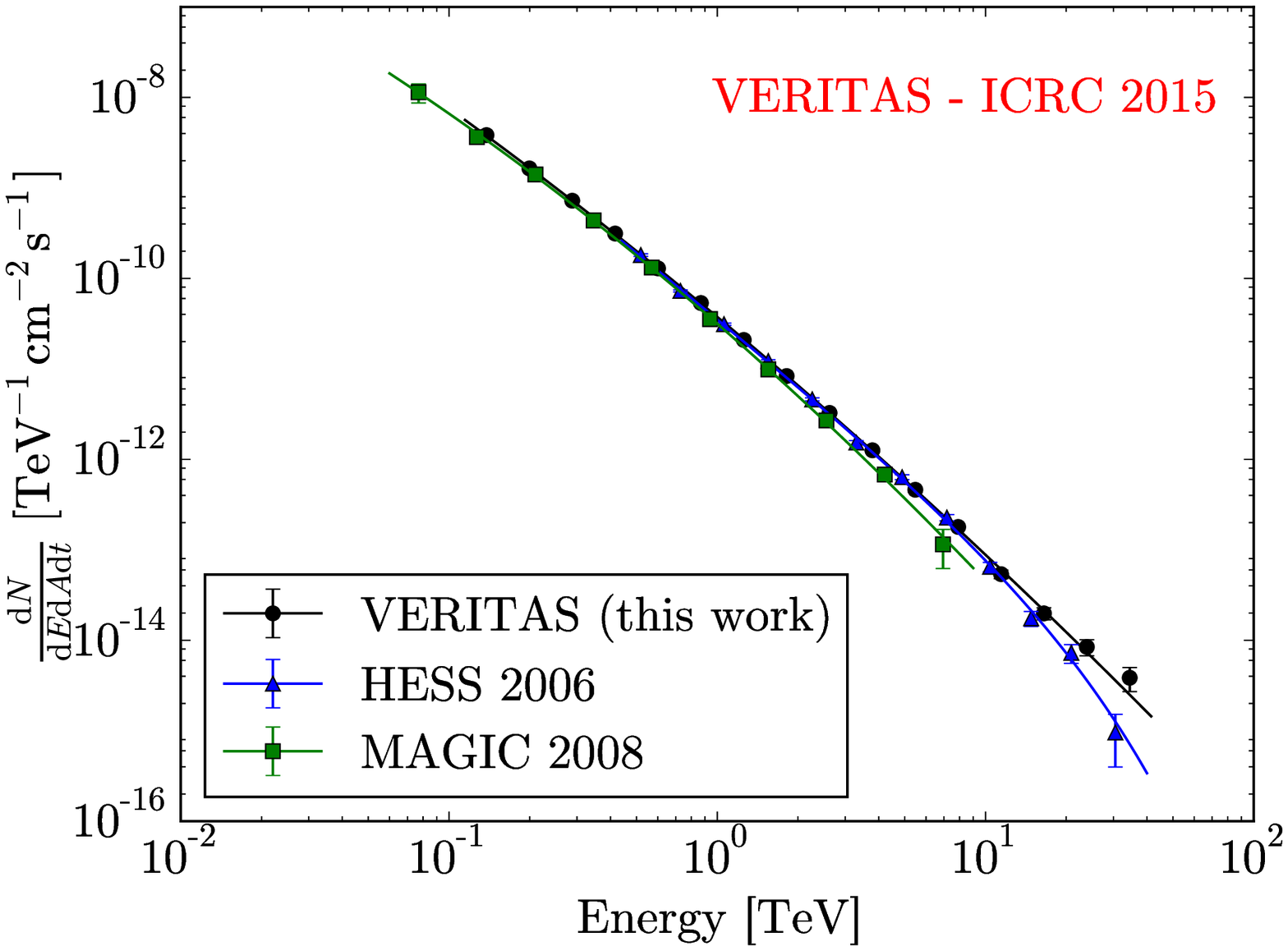}%
  \includegraphics[width=0.5\textwidth]{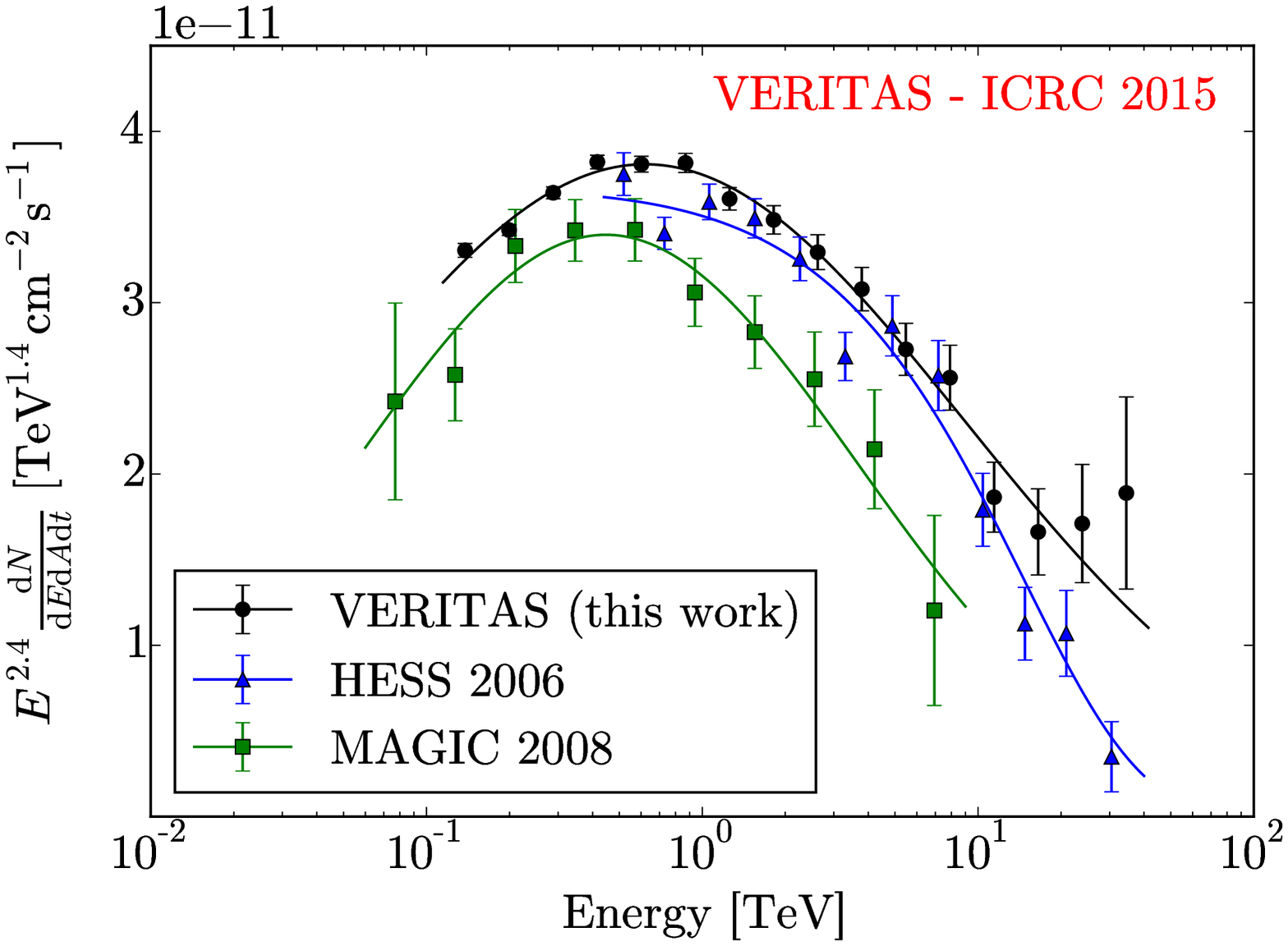}
  \caption{
      \label{fig:spectrum}
      Left panel: The differential energy spectrum of the Crab Nebula. Right plot: The differential energy spectrum multiplied with $E^{2.4}$.  Data from HESS\cite{HESS2006} and MAGIC\cite{MAGIC2008} are shown for comparison. Error bars represent statistical error only.
  }
\end{figure}

The reconstructed spectrum is shown in Figure \ref{fig:spectrum}. The spectrum fits well to a log-parabola function $\frac{\mathrm{d} N}{\mathrm{d} E\mathrm{d} A\mathrm{d} t} = f_{0}\left(\frac{E}{E_{0}}\right)^{-\alpha+\beta\log(E/E_{0})}$. The fit gives a normalization of $f_0 = (3.75 \pm 0.03 )\times 10^{-11}$ $\mathrm{TeV^{-1} cm^{-2} s^{-1}}$, a photon index $\alpha = -2.467 \pm 0.006$, and a curvature parameter $\beta = -0.16 \pm 0.01 $. The fit is performed in the energy range of 115 GeV to 42 TeV, the quality of the fit is $\chi^2 / \mathrm{ndof} = 12.9 / 13$. Only statistical errors are considered here, detailed systematic studies are ongoing. Extensive studies have been perform on subsets of the data. Only small systematic shifts in the spectrum are observed over different cut levels, years, months of the year, wobble offsets, and elevation.

\section{Morphology}
\label{morphology}
The morphology of the Crab Nebula is not yet resolved in gamma rays. It is expected that the size of the inverse-Compton nebula is confined to the standing reverse shock and that its size decreases with increasing gamma-ray energy \citep{deJagerHarding1992}. At 100\,GeV the expected size of the nebula is about $1\farcm5$ \citep{deJagerHarding1992}. From observations the extension has been constrained to be less than $2\farcm4$ at 250\,GeV \citep{MAGIC2008} and less than  $2\farcm0$ above 1\,TeV \citep{HEGRA2000}. 

With VERITAS, like with other gamma-ray experiments, it is not possible to resolve arcminute sized structures. Event by event angular resolution of VERITAS is approximately $4\arcmin$ at 100\,GeV. A source with an extension of $1\arcmin$, results in a subtle broadening of the measured spatial distribution of excess events. In the present case where a radial extent of the Crab Nebula of about $1\farcm5$ is expected the spatial distribution would be 5\% broader than for a true point source.

To perform this analysis, the uncertainty of the point spread function of the telescope must be know to better than the expected broadening of the spatial distribution. Monte Carlo simulation of the detector proved to be in adequate as it was dominated by systematic errors. Measuring the point spread function by observing sources far in the sky from the Crab, such as Markarian 421 and Markarian 501, are also inadequate due to the geomagnetic effect. The blazar VER~J0521+211\citep{VERJ0521}, located only 3 degrees away from the Crab was selected as an template to represent the instruments point spread function. 

The search for an extension of the Crab Nebula is performed in the parameter  $\theta^2$, which measures the radial distance squared between the reconstructed event origin in the sky and the location of the gamma-ray source. For the calculation of $\theta^2$ it is assumed that the maximum of the spatial distribution of excess events coincides with the position of the Crab pulsar in radio: RA = 83.63308, Dec = 22.01450 (J2000)\citep{CrabRadioPosition} in case of the Crab Nebula data and that for VER~J0521+211 the position of the maximum coincides with its optical counterpart RGB~J0521.8+2112: RA = 80.441538, Dec = 21.214273 (J2000)\citep{optialCounterpart}.

The $\theta^2$-distributions for each data set are fitted with a hyperbolic-secant function by means of an unbanned maximum-likelihood method. The radial component of a two dimensional hyperbolic-secant function
\begin{equation}
S(\theta^2;w) = \frac{2 N}{\pi w}\cdot \mbox{sech}(\sqrt{\theta^2}/w)
\end{equation}
is a good empirical analytical description of the leptokurtic $\theta^2$-distribution after background subtraction. 
The parameters of $S$ are $\theta^2$ and the width of the distribution $w$. $w$ denotes the 55.1\% containment radius.

Because the camera acceptance is not uniform, the background as a function of $\theta^2$ is not flat. When restricting the fit to within $1^\circ$ of the fit the background can be characterized by a linear function. This function is fit by taking background data from the half of the field of view opposite from the source and scrabbling it about the center of the camera and fitting the distribution of $\theta^2$ to a first order polynomial.

For the actual fit of the excess distribution the following two-dimensional likelihood function is defined:
\begin{equation}
f(n_{sig},w) = -2 \sum_{\theta^2} \log\left[ \frac{n_{sig}}{n_{total}} S(\theta^2;w) + \left(1-\frac{n_{sig}}{n_{total}}\right) B( \theta^2 ) \right]
\end{equation}
where the summation is over all events, $n_{sig}$, the number of signal events, is used as a nuisance parameter, and $n_{total}$ is the total number of reconstructed events after event selection. The likelihood function is minimized in the range of $0^\circ<\theta<1^\circ$ with the \textsc{Minuit} software package.

\begin{figure}[tb]
  \centering
  \includegraphics[width=0.5\textwidth]{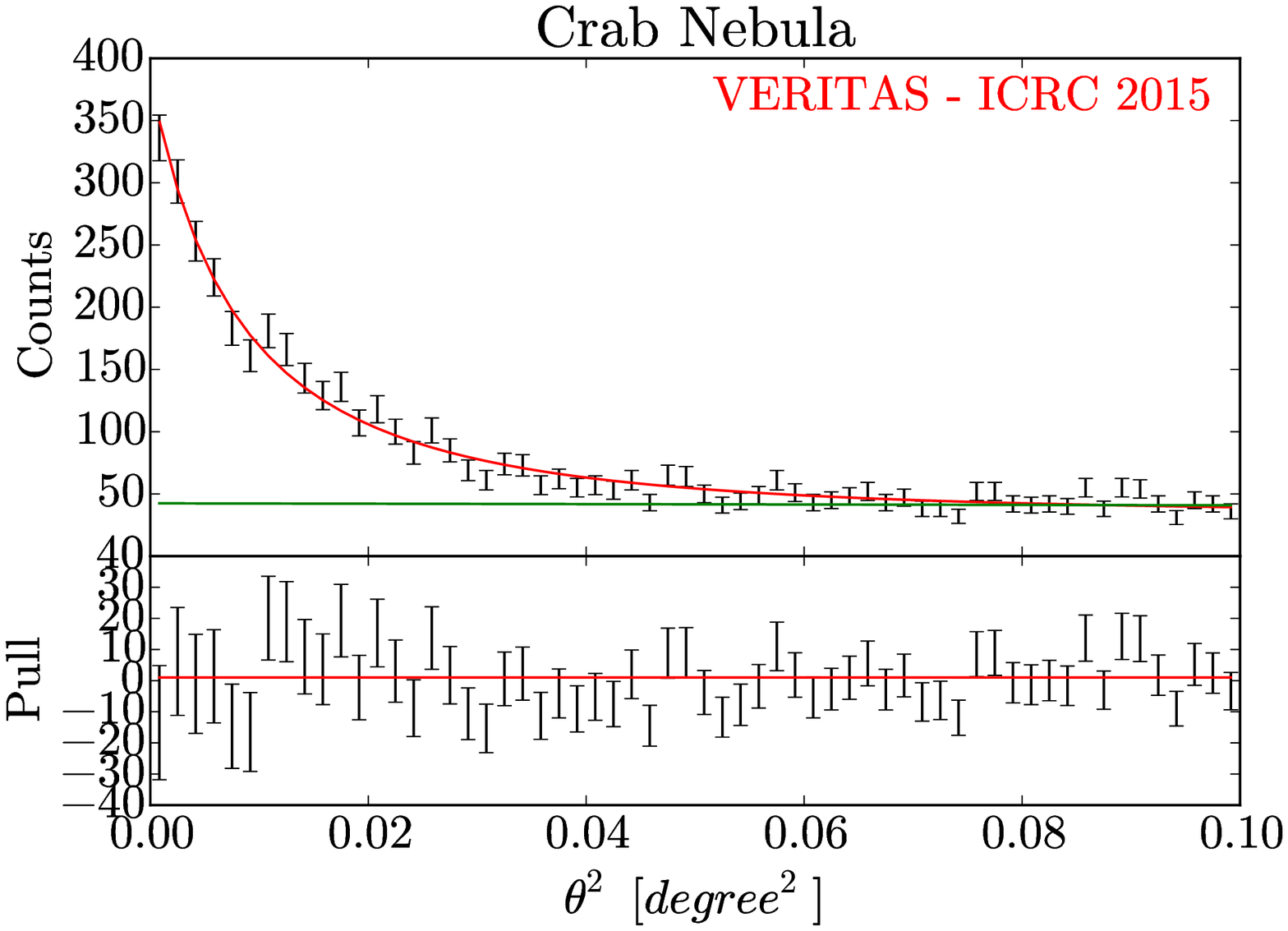}%
  \includegraphics[width=0.5\textwidth]{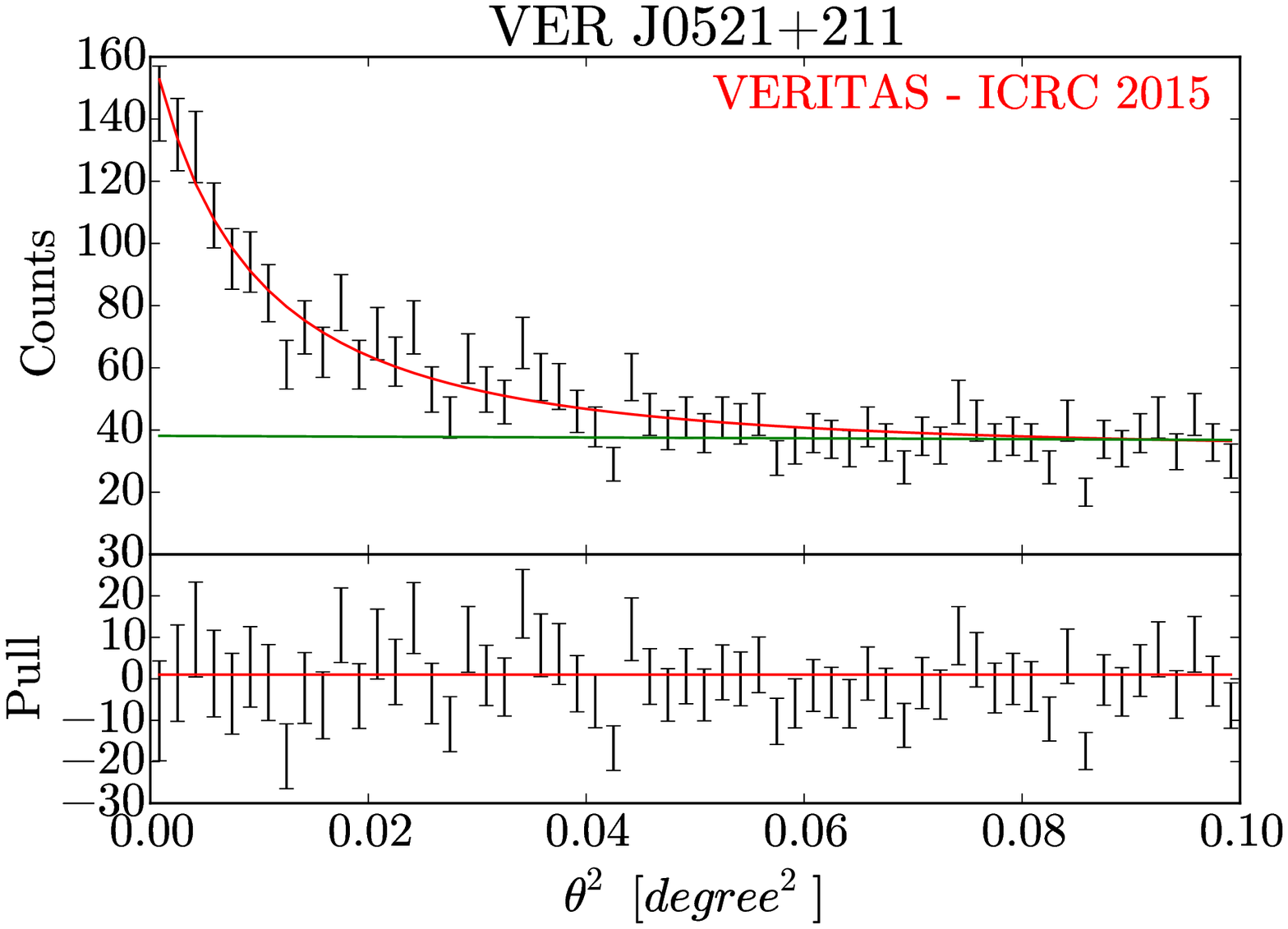}
  \caption{
    \label{thetaSquareDistributionWithRelativeResiduals}
    \emph{Left:} The results of an unbinned maximum likelihood fit of the extent of the Crab Nebula are shown in red. The fit was performed using events with reconstructed energies between 160 and 200\,GeV. Binned data is shown in black for comparison. The Background only hypothesis is shown in green. Lower plot shows the residuals. \emph{Right:} Same plot is shown for the VER~J0521+211 which is used as a template to measure the point spread function.
  }
\end{figure}

\begin{figure}[tb]
  \centering
  \includegraphics[width=0.8\textwidth]{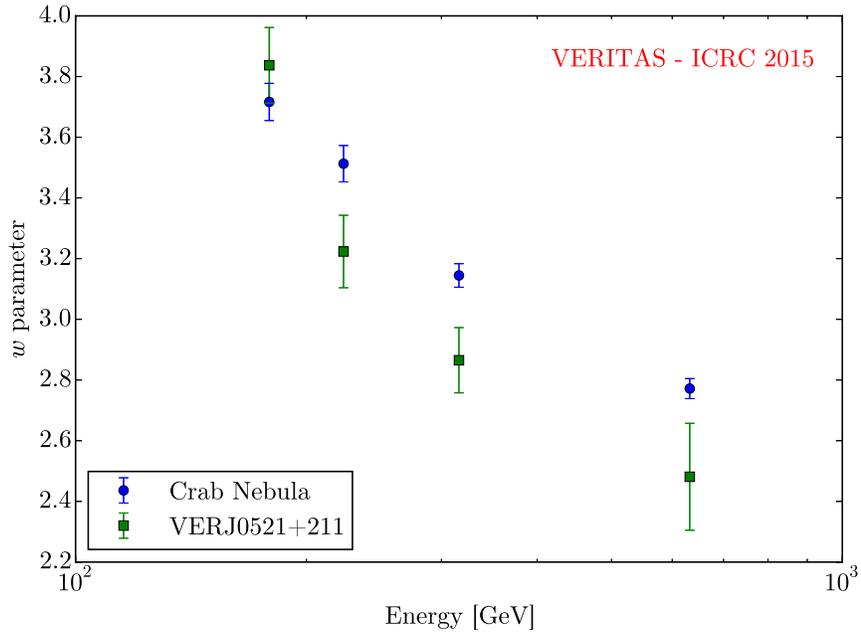}
  \caption{
    \label{w_param}
    The measured width parameter as a function of energy, for both the Crab Nebula and the blazar VER~J0521+211. The lowest energy point is used to set a limit on the extent of the Crab Nebula.
  }
\end{figure}

Figure \ref{thetaSquareDistributionWithRelativeResiduals} shows the $\theta^2$-distribution of the gamma-ray signal of the Crab Nebula including the best fit result averaged over 20 iterations. Figure \ref{w_param} shows the results of the energy fit in four different energy bins.


Source extension is not evident in the data, so an upper limit on the size of the nebula is determined with the unified method by Feldman and Cousins \citep{FeldmanCousins1998}. For the purpose of constructing confidence intervals, pseudo-experiments are performed to empirically determine the probability distribution function $p(w;s)$ where $s$ is the source size. For each source size $10^5$ pseudo-experiments or trials are executed. In each trial the origins of the $N_{signal}$ events are drawn from a normal distribution in both $x$ and $y$ with $\sigma=s$. 

 The uncertainty in the PSF is included in the limit by determining the PSF for each trial from the normal distribution $N(w^{J0521},(\sigma_w^{J0521})^2)$. Each event in a trial is given an offset from its source location randomly chosen from the radial sech function with the PSF for the pseudo-experiment. Background events are added by taking events from the Crab data after adding a random rotation with respect to the center of the camera. These distributions for the Crab Nebula and VER~J0521+211 are then fit using the same unbinned maximum likelihood fitting code as described above.

For the study of the nebula extension the analysis is restricted to post-camera-upgrade data. Including more Crab Nebula data in the analysis does not impact the results presented here due to the dominance of statistical uncertainties in the VER~J0521+211 data. Exact values for upper limits are still in progress.

\section{Variability}
\label{variability}

\begin{figure}
    \includegraphics[width=1.0\textwidth]{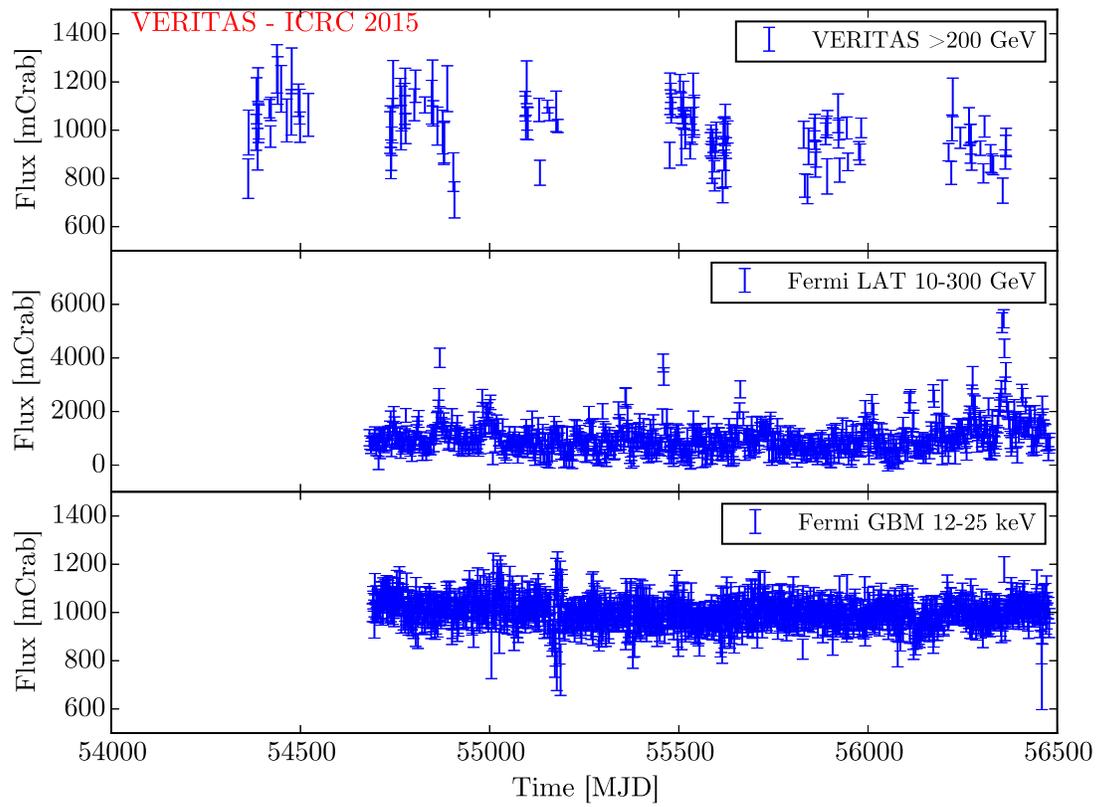}
    \caption{Light curves of the Crab Nebula as measured by VERITAS, \emph{Fermi}-LAT, and \emph{Fermi}-GBM. All light curves are normalized to  crab units where 1 crab unit is equal to the average of the light curve.}
    \label{fig:lightcurve}
\end{figure}

\begin{figure}
  \includegraphics[width=0.5\textwidth]{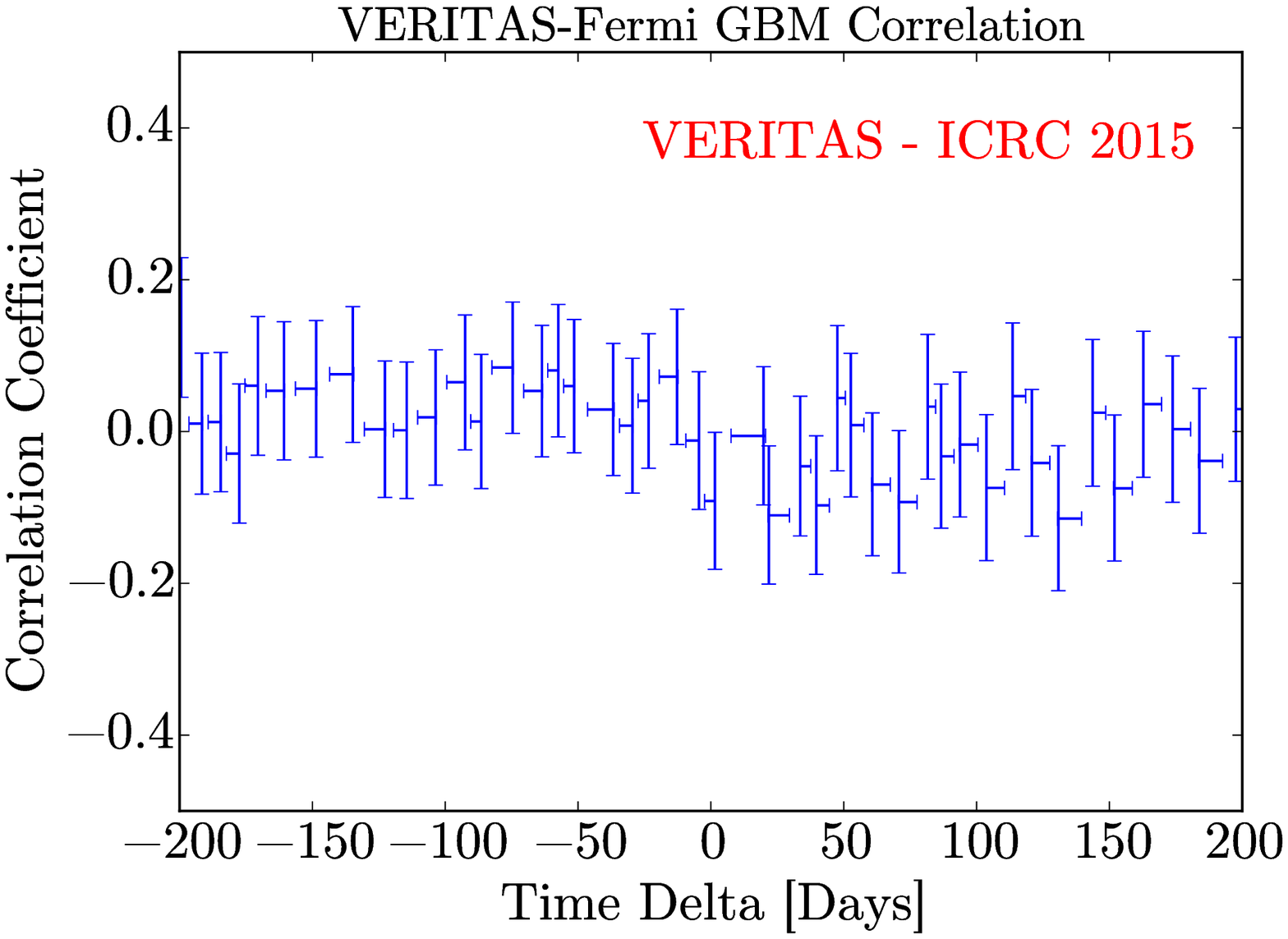}
  \includegraphics[width=0.5\textwidth]{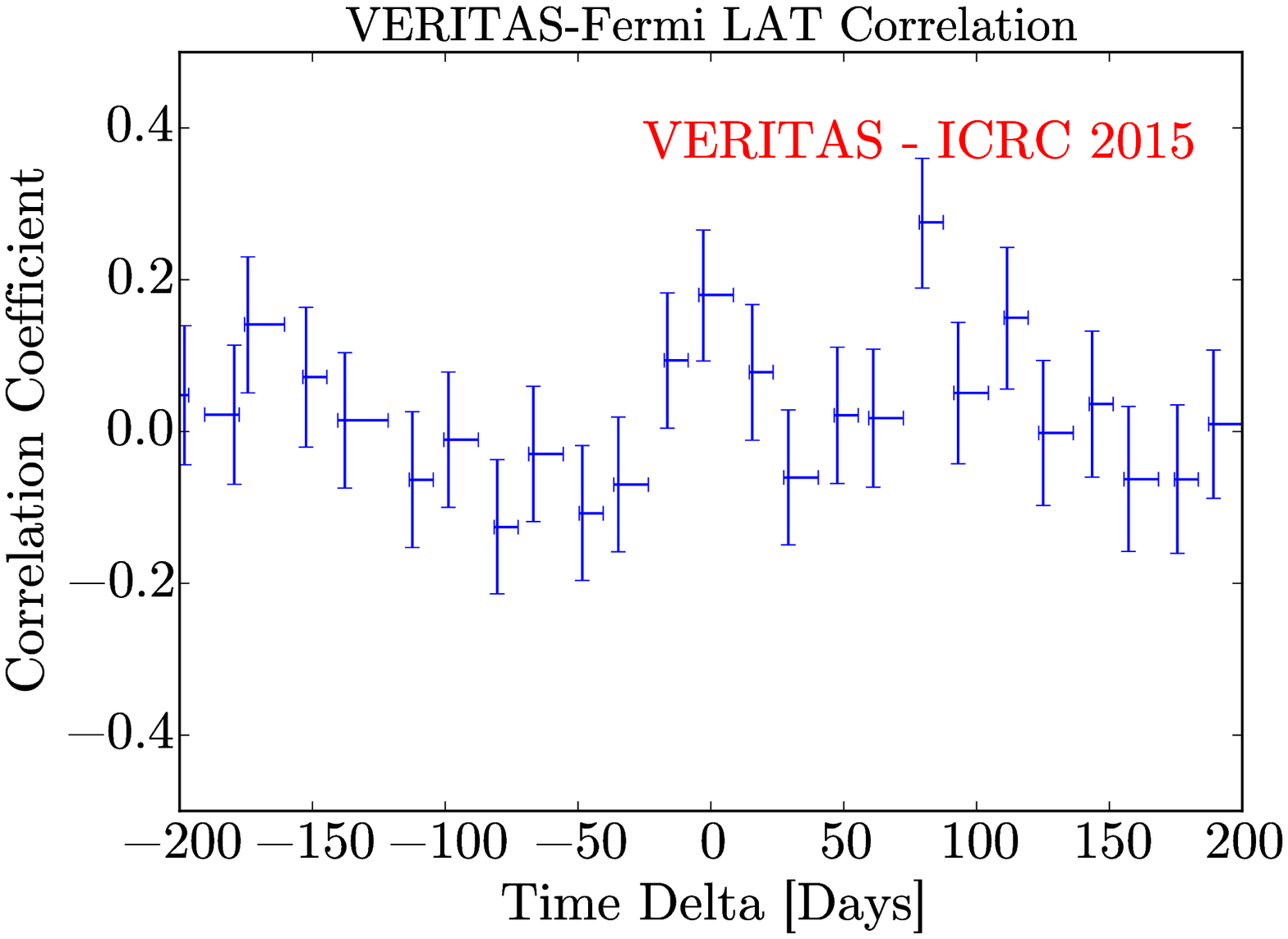}
  \caption{\emph{Left:} Correlation coefficients calculated as a function of a delay between the VERITAS light curve and the \emph{Fermi}-GBM light curve for the Crab Nebula. \emph{Right:} Same figure for a correlation search between VERITAS and \emph{Fermi}-LAT light curves}
  \label{fig:correlations}
\end{figure}

To investigate the variability of the Crab Nebula we calculated the average flux above 200\,GeV in bins of one day for every night which VERITAS observed the Crab Nebula. The light curve is shown in figure \ref{fig:lightcurve}. Observations where the elevation was too low to observe the source at an energy of 200\,GeV were not used in this analysis. No strong evidence of variability is seen in the data. 
We also investigate variability of the flux by comparing the VHE light curve to the one observed in the hard x-ray band by \emph{Fermi}-GBM\cite{StandardCandleFlickers}, and in the high-energy gamma-ray band by \emph{Fermi}-LAT. The light curves for these instruments are shown in Figure \ref{fig:lightcurve}. The correlation between two light curves was calculated using the z-transformed discrete correlation function (ZDCF), which is widely used to estimate the cross-correlation function of discrete sparse astronomical time series\cite{zdcf}. ZDCF will calculate correlation coefficients for all time scales up to the duration of the observation. Figure \ref{fig:correlations} shows the results of this analysis. No evidence for correlated variability is seen between any of the energy bands on any time scale.
The results represented here are consistent with an analysis performed by VERITAS searching for correlation specifically during a flare of the Crab Nebula in 2013\citep{VERITAS2014}.

\section{Discussion}
In this proceedings, we have shown the results of six years of observations of the Crab Nebula with VERITAS. The spectrum is consistent within quoted systematic uncertainties with the spectrum published with other instruments. We investigated the morphology of the source and saw no evidence of an extension. The lightcurve analysis shows no correlation with variability of the source in other wavebands.  Analysis of these data including more detailed treatments of systematic errors are still ongoing. 

\section*{Acknowledgements}
This research is supported by grants from the U.S. Department of Energy Office of Science, the U.S. National Science Foundation and the Smithsonian Institution, and by NSERC in Canada. We acknowledge the excellent work of the technical support staff at the Fred Lawrence Whipple Observatory and at the collaborating institutions in the construction and operation of the instrument. The VERITAS Collaboration is grateful to Trevor Weekes for his seminal contributions and leadership in the field of VHE gamma-ray astrophysics, which made this study possible.

\bibliographystyle{JHEP}
\bibliography{crab}

\end{document}